\documentclass[twocolumn,aps,showpacs,floatfix,prc]{revtex4}
\usepackage{graphicx}
\usepackage[dvips]{epsfig}

\begin{document}

\title{ Rossby modes in neutron stars as sources of gravitational waves }

\author{ Sujan Roy$^{1*\S}$, Somnath Mukhopadhyay$^{2*\S}$, Joydev Lahiri$^{3*}$, Debasis Atta$^{4\dagger}$, Partha Roy Chowdhury$^{5\ddagger}$  and D. N. Basu$^{6*\S}$ }

\affiliation{$^*$ Variable  Energy  Cyclotron Centre, 1/AF Bidhan Nagar, Kolkata 700064, India}
\affiliation{ $^{\dagger}$ Government General Degree College, Kharagpur II, West Bengal 721149, India}
\affiliation{ $^{\ddagger}$ Chandrakona Vidyasagar Mahavidyalaya, Chandrakona Town, Paschim Medinipur, West Bengal 721201, India}
\affiliation{$^{\S}$ Homi Bhabha National Institute, Training School Complex, Anushakti Nagar, Mumbai 400085}

\email[E-mail 2: ]{sujan.kr@vecc.gov.in}
\email[E-mail 2: ]{somnathm@vecc.gov.in}
\email[E-mail 3: ]{joy@vecc.gov.in}
\email[E-mail 4: ]{debasisa906@gmail.com}
\email[E-mail 5: ]{royc.partha@gmail.com}
\email[E-mail 6: ]{dnb@vecc.gov.in}

\date{\today }

\begin{abstract}

    In the present work, we explore the Rossby mode instabilities in neutron stars as sources of gravitational waves. The intensity and time evolution of the emitted gravitational waves in terms of the amplitude of the strain tensor are estimated in the slow rotation approximation using $\beta$-equilibrated neutron star matter obtained from density dependent M3Y effective interaction. For a wide range of neutron star masses, the fiducial gravitational and various viscous time scales, the critical frequencies and the time evolutions of the frequencies are calculated. The dissipative mechanism of the Rossby modes is considered to be driven by the shear viscosity along the boundary layer of the solid crust-liquid core interface as well as in the core and the bulk viscosity. It is found that neutron stars with slower frequency of rotation, for the same mass, radius and surface temperature, are expected to emit gravitational waves of higher intensity.     

\vspace{0.2cm}    

\end{abstract}


\maketitle

\noindent
\section{Introduction}
\label{Section 1} 

 Neutron stars are formed in supernova explosions by the collapse of the core of a star which is near the end point of thermonuclear evolution. When a neutron star forms, it pulsates wildly. The initial energy of pulsation will be of the order of the kinetic energy of collapse, which is between $\sim$ 0.01 and $\sim$ 0.2 of the rest mass-energy of the neutron star. The gradual transfer of this huge pulsation energy to the supernova envelope which surrounds the star may have important observational consequences. To explore these one needs to analyse radial and non-radial pulsations of relativistic stellar models. Particularly the non-radial perturbation analysis of super-massive star and neutron star (NS) is vital because of its intimate connection to gravitational pulsation of star. The corresponding General Relativistic analysis of non-radial perturbation over hydrostatic equilibrium using Regge-Wheeler choice of gauge for non-rotating or slowly rotating spherically symmetric star can be found in Ref.\cite{Thorne1967}. As the emitted gravitational wave (GW) radiation caused by non-radial perturbation carries energy it should cause damping of the perturbation. Depending upon various characteristics these non-radial perturbations are recognized by different names. Among various modes, the Rossby mode (r-mode) non-radial perturbation in case of highly dense object like NS is of fundamental importance because of its special feature that the flow pattern is in the opposite direction to the star's rotation as seen by an observer at rest on the star (retrograde) and in the same direction to star's rotation when observed by an observer at infinity (prograde). Interestingly Chandrasekhar-Friedman-Schutz (CFS) showed that gravitational radiation emitting due to any perturbation which is retrograde in co-rotating frame and progarde in inertial frame will hasten the perturbation further instead of damping it. This phenomenon clearly indicates that gravitational radiation drives r-modes towards instability by causing their amplitudes to grow exponentially provided the damping mechanisms are not significant enough. In the present work various viscous damping mechanisms such as shear and bulk viscosities have been considered, although possibility of other non-linear damping mechanisms should not be ignored as will be explained in Section-IV. Thus the stability of r-mode is determined by estimating which term is dominating. 
 
    The instability in the mode grows because of gravitational wave emission which is opposed by the viscosity \cite{Lindblom1987}. For the instability to be relevant, it must grow faster than it is damped out by the viscosity. Therefore, the time scale for gravitationally driven instability needs to be sufficiently short compared to the viscous damping time scale. The amplitude of r-modes evolves with time dependence $ e^{i\omega t - t/\tau } $ as a consequence of ordinary hydrodynamics and the influence of the various dissipative processes. The imaginary part of the frequency $1/\tau$ is determined by the effects of gravitational radiation, viscosity, etc \cite{Lindblom2000,Lindblom1998,Owen1998}. The time scale associated with the different process involve the actual physical parameters of the neutron star. In computing these physical parameters the role of nuclear physics comes into picture, where one gets a platform to constrain the uncertainties existing in the nuclear Equation of State (EoS). The present knowledge on nuclear EoS under highly isospin asymmetric dense situation is quite uncertain. So correlating the predictions of the EoSs obtained under systematic variation of the physical properties, to the r-mode observables can be of help in constraining the uncertainty associated with the EoS. For the present illustrative calculations, the $\beta$-equilibrated NS matter EoS obtained from the density dependent M3Y (DDM3Y) effective nucleon-nucleon interaction \cite{BCS08} has been employed.  
 
    The rate of change in perturbation energy due to GW for non-radial perturbation like r-mode is $\omega\left(\omega+l\Omega\right)\sum N_{l}\omega^{2l}\left(\mid\delta D_{ll}\mid^2+\mid\delta J_{ll}\mid^2\right)$ for $l\geq 2$ with
$\delta D_{ll}$, $\delta J_{ll}$ are mass and current multipole moments of the perturbation respectively \cite{Lindblom1998}. The quantity $\omega$ being the angular frequency of the perturbation connected to the angular frequency of the star $\Omega $ through 

\begin{equation}
\omega=-\frac{\left(l-1\right)\left(l+2\right)}{l+1} \Omega.
\label{eq1}
\end{equation}
The rate of change of perturbation energy due to GW emission alludes the perturbation to be increasing with increasing $\Omega$. So the rapidly rotating NSs are expected to emit considerable amount of gravitational radiation which carries the energy and angular momentum of the star thereby providing possible elucidation of spin down mechanism of hot NSs as will be elaborated in Section-II. It is important to mention that the General Relativistic treatment of perturbation in NS matter reveals that spherical harmonic part of the solution exists only in the case of $l=m$ for $l\geq 2$. Among all the perturbation modes, $l=2$ is particularly important because for each increment in $l$ the gravitational pumping decreases by almost one order of magnitude whereas the damping increases by $\sim 20\%$ \cite{Lindblom1998} making other modes stable against gravitational emission. Hence, considering $l=2$ the r-mode instability window has been calculated and recent data are verified with respect to that window. As for this particular r-mode perturbation emitted gravitational radiation yields feeble strain tensor amplitude $h_0$ to be of the order of $10^{-28}$ (will be discussed briefly in Section-V) which lies below anticipated threshold for detection of Advanced LIGO \cite{Mahmoodifar2013} and the angular momentum secession too is expected to be quite small for slowly rotating NS. Gradually $l=2$ r-mode perturbation should be resulting in considerably slow angular frequency evolution. The angular frequency evolution and the rate of carrying angular momentum by GW in such case for different star configurations have been calculated in the present work and will be described in the Section-V.
     
\noindent
\section{Spin down due to r-mode instability} 
\label{Section 2}

    In the present work, the evolution of the r-mode due to the competition of gravitational radiation and dissipative impact of viscosities has been studied. To achieve this objective it is necessary to survey the consequences of radiation and dissipative mechanisms on the evolution of mode energy which can be expressed as the integral of the non-radial fluid perturbation \cite{Lindblom1998,Lindblom1999},
    
\begin{equation}
\widetilde{E}=\frac{1}{2}\int{\left[ \rho \delta \vec{v}.\delta \vec{v}^{*}+\left(\frac{\delta p}{\rho}-\delta \Phi \right)\delta \rho^{*}\right]}d^{3}r,
\label{eq4}
\end{equation}
where $\rho$ is the mass density profile of the star, $\delta \vec{v}$, $\delta p$, $\delta \Phi$ and $\delta \rho $ are perturbations in the velocity, pressure, gravitational potential and density due to oscillation of the mode, respectively. Various time scales of an r-mode \cite{Lindblom1998} is expressed as 

\begin{equation}
\frac{1}{\tau_{i}}=-\frac{1}{2\widetilde{E}}\left(\frac{d\widetilde{E}}{dt}\right)_{i},
\label{eq5}
\end{equation}
where the index `$i$' refers to the gravitational radiation, bulk and shear viscosities in the fluid core and viscous dissipation at the boundary layer between the crust and the core. The expressions of the terms involved in the current and mass multipoles are deduced in Refs.\cite{Ipser1991,Thorne1980}. 
    
    The expression of energy of the r-mode in Eq.(\ref{eq4}) can be reduced to an one-dimensional integral \cite{Lindblom1998,Vidana2012} for the lowest order expressions of $\delta \vec{v}$ and $\delta \rho$ in the small angular velocity limit as

\begin{equation}
\widetilde{E}=\frac{1}{2}\alpha_r^{2} R^{-2l+2} \Omega^{2} \int^{R}_{0} \rho(r) r^{2l+2} dr, 
\label{eq6}
\end{equation}
where $\alpha_r$ is the dimensionless amplitude of the mode, $R$ is the radius, $\Omega$ is the angular velocity and $\rho(r)$ is the radial dependence of the mass density of NS. The fact that the expressions of $(\frac{d\widetilde{E}}{dt})$ due to gravitational radiation \cite{Thorne1980,Owen1998} and viscosities \cite {Ipser1991,Owen1998,Lindblom2000} are known, Eq.(\ref{eq5}) can be used to evaluate the imaginary part $\frac{1}{\tau}$.

    When the angular velocity of a NS goes beyond the critical value $\Omega_c$, the instability of mode sets in and the star emits gravitational radiation which takes away the energy and angular momentum, resulting the star to spin down to the region of stability. The evolution of the angular velocity \cite{Owen1998}, as the angular momentum is radiated to infinity by gravitational radiation, is given by
    
\begin{equation}
\frac{d\Omega}{dt}=\frac{2\Omega}{\tau_{GR}}\frac{\alpha_r^2Q}{1-\alpha_r^2Q},
\label{eq7}
\end{equation}
where $\tau_{GR}$ is the gravitational time scale, $\alpha_r$ is the dimensionless r-mode amplitude and $Q=3 \widetilde{J}/2 \widetilde{I}$ with

\begin{equation}
\widetilde{J}=\frac{1}{MR^4}\int^{R}_{0}\rho(r)r^{6} dr
\label{eq8}
\end{equation}
and
\begin{equation}
\widetilde{I}=\frac{8\pi}{3MR^2}\int^{R}_{0}\rho(r)r^{4} dr.
\label{eq9}
\end{equation}
From saturation using linear dissipations, spin equilibrium and thermal equilibrium \cite{Mahmoodifar2013}, $\alpha_r$ can be estimated within a wide range of values from $1-10^{-8}$. Under the ideal consideration that the heat generated by the shear viscosity is same as that taken out by the emission of neutrinos and photons \cite{Bondarescu2009,Moustakidis2015}, Eq.(\ref{eq7}) can be solved easily \cite{Mu18} for constant $\alpha_r$ \cite{Moustakidis2015} as $\Omega(t)=\left(\Omega^{-6}_{in}-\textsl{C}t\right)^{-1/6}$ where $\textsl{C}=\frac{12\alpha_r^2Q}{\widetilde{\tau}_{GR}\left(1-\alpha_r^2Q\right)}\frac{1}{\Omega_0^6}$ and $\Omega_{in}$ is considered as a free parameter whose value corresponds to be the initial angular velocity. The spin down rate can be obtained from Eq.(\ref{eq7}) to be $\frac{d\Omega}{dt}=\frac{\textsl{C}}{6}\left(\Omega^{-6}_{in}-\textsl{C}t\right)^{-7/6}$. The neutron star spin shall decrease continually until it approaches its critical angular velocity $\Omega_c$. The time $t_c$ taken by neutron star to evolve from its initial value $\Omega_{in}$ to its minimum value  $\Omega_{c}$ is given by $t_c=\frac{1}{\textsl{C}}\left(\Omega_{in}^{-6}-\Omega_{c}^{-6}\right)$.

\noindent
\section{Bulk and shear viscosities} 
\label{Section 3}

    The time scale $\tau$ of the imaginary part of the r-mode oscillation can be decomposed into the sum of the contributions of all the different dissipative processes as 

\begin{equation}
\frac{1}{\tau(\Omega,T)}=\frac{1}{\tau_{GR}(\Omega)}+\frac{1}{\tau_{BV}(\Omega,T)}+\frac{1}{\tau_{SV}(T)}+\frac{1}{\tau_{VE}(\Omega,T)},
\label{eq10}
\end{equation}
where $1/\tau_{GR}$, $1/\tau_{BV}$, $1/\tau_{SV}$ and $1/\tau_{VE}$ are the contributions from gravitational radiation, bulk and shear viscous time scales in the fluid core and viscous dissipation in the crust-core boundary layer, respectively, and are given by \cite{Owen1998,Lindblom2000}

\begin{eqnarray}
\frac{1}{\tau_{GR}}=-\frac{32 \pi G \Omega^{2l+2}}{c^{2l+3}} \frac{(l-1)^{2l}}{[(2l+1)!!]^2}\left(\frac{l+2}{l+1}\right)^{(2l+2)}\nonumber \\
 \times\int^{R_{c}}_{0}\rho(r)r^{2l+2} dr, 
\label{eq11}
\end{eqnarray}

\begin{eqnarray}
\frac{1}{\tau_{BV}}&=&\frac{4\pi R^{2l-2}}{690}\left(\frac{\Omega}{\Omega_0}\right)^{4}\left(\int^{R_c}_{0}\rho(r)r^{2l+2} 
dr\right)^{-1}\nonumber \\ 
&&\times \int^{R_c}_{0}\xi_{BV}\left(\frac{r}{R}\right)^{6}\left[1+0.86\left(\frac{r}{R}\right)^{2}\right] r^{2} dr \hspace{0.5cm}
\label{eq12}
\end{eqnarray}
where $\xi_{BV}$ is the bulk viscosity, the shear viscous dissipation time scale $1/\tau_{SV}$ in the fluid core is given by \cite{Lindblom1998} 
\begin{equation}
\frac{1}{\tau_{SV}}=(l-1) (2l+1) \left(\int^{R_c}_{0}\rho(r)r^{2l+2} dr\right)^{-1}\int^{R_c}_{0}\eta_{SV} r^{2l} dr 
\label{eq13}
\end{equation}
where $\eta_{SV}$ is the shear viscosity, and for the viscous dissipation in the crust-core boundary layer

\begin{eqnarray}
\frac{1}{\tau_{VE}}=\left[\frac{1}{2\Omega} \frac{2^{l+3/2}(l+1)!}{l(2l+1)!!I_{l}}\sqrt{\frac{2\Omega R_{c}^{2} \rho_{c}}{\eta_c}}\right]^{-1}\nonumber \\
\times\left[\int^{R_{c}}_{0} \frac{\rho(r)}{\rho_{c}}\left(\frac{r}{R_{c}}\right)^{2l+2} \frac{dr}{R_c}\right]^{-1}, 
\label{eq14}
\end{eqnarray}
where $G$ and $c$ are the gravitational constant and velocity of light respectively; $R_{c}$, $\rho_{c}$, $\eta_{c}$ in Eq.(\ref{eq14}) are the radius, density and shear viscosity of the fluid at the outer edge of the core respectively. 

    The paramount importance of the viscous dissipation in the crust–core boundary layer was shown first by Bildsten and Urshomirsky  \cite{Bildsten2000}. The shear viscosity time scale in Eq.(\ref{eq14}) is obtained by considering the shear dissipation in the viscous boundary layer between solid crust and the liquid core with the assumption that the crust is rigid and hence static in rotating frame \cite{Lindblom2000}. The motion of the crust due to mechanical coupling to the core effectively increases $\tau_{VE}$ by $(\frac{ \Delta v}{v})^{-2}$, where $\frac{\Delta v}{v}$ is the difference in the velocities in the inner edge of the crust and outer edge of the core divided by the velocity of the core \cite{Levin2001}. Bildsten and Ushomirsky  \cite{Bildsten2000} have first estimated this effect of solid crust on r-mode instability and shown that the shear dissipation in this viscous boundary layer decreases the viscous damping time scale by more than $10^5$ in old accreting neutron stars and more than $10^7$ in hot, young neutron stars. $I_{l}$ in Eq.(\ref{eq14}) has the value $I_{2}=0.80411$, for $l=2$ \cite{Lindblom2000}. 

    The bulk viscous dissipation is not significant for temperature of the star below $10^{10}$ K and in this range of temperature the shear viscosity is the dominant dissipative mechanism. However, beyond this temperature range the bulk viscous dissipative mechanism takes over. In the present work, we have studied the instability for temperatures up to $10^{11}$ K. In the temperature range $T \geq 10^9$ K, the dominant contribution to shear is from neutron-neutron (nn) scattering and below $T \leq 10^9$, it is the electron-electron (ee) scattering that contributes to shear primarily \cite{Lindblom2000}. Therefore,
    
\begin{equation}
\frac{1}{\tau_{V}}=\frac{1}{\tau_{V}^{ee}}+\frac{1}{\tau_{V}^{nn}},
\label{eq15}
\end{equation}
where $\tau_V^{ee}$ and $\tau_V^{nn}$ can be obtained from Eq.(\ref{eq13}) when the subscript $V$ stands for $SV$ and from Eq.(\ref{eq14}) when the subscript $V$ stands for $VE$ using the corresponding values of $\eta_{V}^{ee}$ and $\eta_{V}^{nn}$ from the expressions given by \cite{Flowers1979,Cutler1987}

\begin{equation}
\eta_{V}^{ee}=6 \times 10^{6} \rho^{2} T^{-2} ~~~~~({\rm g~cm^{-1}~s^{-1}}), 
\label{eq16}
\end{equation}

\begin{equation}
\eta_{V}^{nn}=347 \rho^{9/4} T^{-2} ~~~~~({\rm g~cm^{-1}~s^{-1}}), 
\label{eq17}
\end{equation}
where all the quantities are given in CGS units and T is measured in K. The bulk viscosity in Eq.(\ref{eq12}) is the dominating dissipation mechanism at high temperature. The bulk viscosity in Eq.(\ref{eq12}) should be determined using $\xi_{BV}$ evaluated consistently for the EoS \cite{Vidana2012}. However, in the present work we have used the approximate expression for $\xi_{BV}$ \cite{Sawyer1989,Lindblom1998,Andersson1999,Andersson2003,Moustakidis2016} expressed in units of ${\rm g~cm^{-1} s^{-1}}$ by  

\begin{equation}
\xi_{BV} =6 \times 10^{-59} \left(\frac{l+1}{2}\right)^{2} \left(\frac{\rm Hz}{\Omega}\right)^{2} \left(\frac{\rho}{{\rm g~cm^{-3}}}\right)^{2} \left(\frac{T}{ K}\right)^{6} 
\label{eq18}
\end{equation} 

    In order to have transparent visualisation of the role of angular velocity and temperature on various time scales, it is useful to factor them out by defining fiducial time scales. Thus, we define fiducial gravitational radiation time scale $\widetilde{\tau}_{GR}$ through the relation \cite{Lindblom2000,Lindblom1998},
    
\begin{equation}
\tau_{GR}=\widetilde{\tau}_{GR} \left(\frac{\Omega_0}{\Omega}\right)^{2l+2},
\label{eq19}
\end{equation}
the fiducial bulk viscous time scale $\widetilde{\tau}_{BV}$ by expression
\begin{equation}
\tau_{BV}=
\widetilde{\tau}_{BV}\left(\frac{\Omega_0}{\Omega}\right)^{2}\left(\frac{10^{11} K}{T}\right)^{6},
\label{eq20}
\end{equation}
the fiducial shear viscous time scale $\widetilde{\tau}_{SV}$ of the core by
\begin{equation}
\tau_{SV}=\widetilde{\tau}_{SV}\left(\frac{T}{10^6 K}\right)^{2} 
\label{eq21}
\end{equation}
and the fiducial shear viscous time scale $\widetilde{\tau}_{VE}$ at junction is defined such that \cite{Lindblom2000,Lindblom1998}

\begin{equation}
\tau_{VE}=\widetilde{\tau}_{VE} \left(\frac{\Omega_0}{\Omega}\right)^{1/2} \left(\frac{T}{10^8 K}\right),
\label{eq22}
\end{equation}
where $\Omega_0=\sqrt{ \pi G \bar{\rho}}$ and $\bar{\rho}= 3M/4 \pi R^3$ is the mean density of NS having mass $M$ and radius $R$. The gravitational radiation tends to drive the $r$-mode to the instability, 
while the viscosity suppresses it. The dissipation effects due to viscosity cause the $r$-mode to decay exponentially as $e^{-t/\tau}$ as long as $\tau > 0$ \cite {Lindblom1998}. In order to make out the role of $\Omega$ and T in various time scales, it is useful to factor them out by defining respective fiducial time scales. The time scale $\tau$ given in the Eq.(\ref {eq10}) can now be expressed as

\begin{eqnarray}
\frac{1}{\tau(\Omega,T)}&=&\frac{1}{\widetilde{\tau}_{GR}}\left(\frac{\Omega}{\Omega_0}\right)^{2l+2}
+\frac{1}{\widetilde{\tau}_{BV}}\left(\frac{\Omega}{\Omega_0}\right)^{2}\left(\frac{T}{10^{11} K}\right)^{6} \nonumber \\
&+&\frac{1}{\widetilde{\tau}_{SV}}\left(\frac{10^6 K}{T}\right)^{2} 
+\frac{1}{\widetilde{\tau}_{VE}}  \left(\frac{10^8 K}{T}\right) \left(\frac{\Omega}{\Omega_0}\right)^{1/2}
\label{eq23}
\end{eqnarray}
where ${\widetilde{\tau}_{GR}}$, ${\widetilde{\tau}_{SV}}$, ${\widetilde{\tau}_{BV}}$ and ${\widetilde{\tau}_{VE}} $ are the respective fiducial time scales that can be defined from Eqs.(\ref{eq11}-\ref{eq14}). At small $\Omega$, the gravitational radiation is small (due to the $\Omega^{2l+2}$ dependence) while the viscosity dominates and keeps the mode stable. But for large angular velocity $\Omega$, the gravitational radiation dominates and drives the mode to the instability. For a given mode $l$, the critical angular velocity $\Omega_c$ is obtained from the condition,

\begin{equation}
\frac{1}{\tau(\Omega_c,T)}=0,
\label{eq24}
\end{equation} 
where $1/\tau$ is given in Eq.(\ref{eq23}). At a given $T$ and mode $l$, the equation for $\Omega_c$ is a polynomial of order $l+1$ in $\Omega_c^2$ and thus each mode has its own characteristic $\Omega_c$ value. Since the smallest mode $l=2$ is the most important one, the study is made for this $l=2$ mode, where the critical frequency is obtained from the solution of Eq.(\ref{eq24}).

\noindent
\section{Limiting the r-mode amplitudes} 
\label{Section4}

    By setting accretion torque in Eq.(12b) of Ref.\cite{Mahmoodifar2013} to zero, the saturation value of $\alpha_r$ can be obtained using $\frac{d\alpha_r}{dt}=0$ which provides very high saturation value of the amplitude $\alpha_r \sim 1$. This would mean extremely high luminosity which has no resemblance with the observations. Thus none of the saturation mechanisms proposed so far can saturate r-modes at low amplitudes required to explain observed luminosities. To constrain the amplitude to reasonable values non-linear dissipative terms needs to be incorporated which are yet to be determined.  Therefore, in this work two different methods for constraining the r-mode amplitude, $\alpha_r$, from observations of LMXB NS transients have been compared. The first one, which gives larger values for $\alpha_r$, is based on the spin equilibrium assumption \cite{Brown2000,Ho2011,Haskell2011,Watts2008} where we assume that in an outburst-quiescence cycle all the spin-up torque due to accretion during the outburst is balanced by the r-mode spin-down torque due to gravitational radiation in the whole cycle. The second one is based on the thermal equilibrium outlined in Ref.\cite{Brown2000}, but rather than estimating the quiescent luminosity using the r-mode amplitude deduced from spin equilibrium, observations of the quiescent luminosity of LMXBs to directly constrain the amplitude of the r-mode have been used.
    
\subsection{Constraints from `Spin Equilibrium'}
    
    This is similar to the prescription considered by previous authors \cite{Brown2000,Ho2011}, but rather than using a ``fiducial'' torque estimated from the long-term average $\dot M$, we can now use the observed spin-up rates and outburst properties to directly constrain the torque. Therefore we have  
    
\begin{equation}
  2\pi I \dot{\nu} \Delta =  \frac{2J_c}{\tau_{GR}}
\label{eq25}
\end{equation} 
\noindent
where $I=MR^2\widetilde{I}$ is the moment of inertia and $J_c$ is the canonical angular momentum of the mode given by

\begin{equation}
  J_c = -\frac{3}{2}\widetilde{J}M R^2 \Omega \alpha_r^2,
\label{eq26}
\end{equation} 
\noindent
the dimensionless quantities $\widetilde{J}$ and $\widetilde{I}$ are provided in Eq.(\ref{eq8}) and Eq.(\ref{eq9}), respectively, $\dot{\nu}$ is the spin-up rate during outburst and $\Delta = (t_o/t_r)$ is the ratio of the outburst duration $t_o$ to the recurrence time $t_r$.

\begin{figure}[t]
\vspace{0.0cm}
\eject\centerline{\epsfig{file=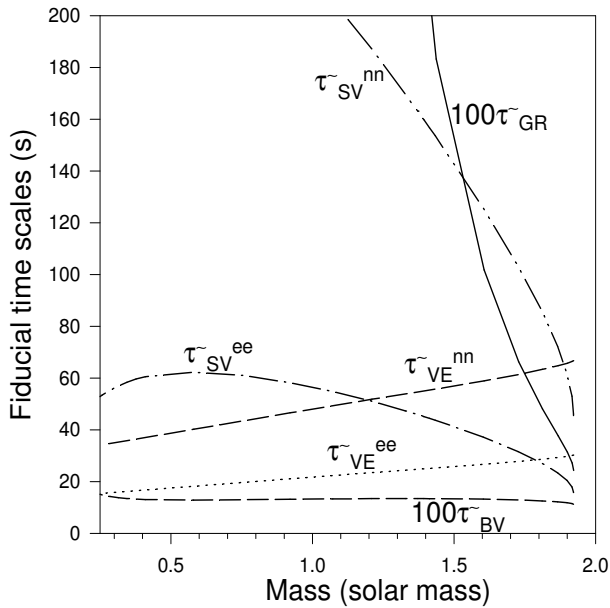,height=8cm,width=8cm}}
\caption{Plots of fiducial timescales with gravitational mass of neutron stars with DDM3Y EoS.} 
\label{fig1}
\vspace{0.0cm}
\end{figure}
         
\subsection{Constraints from `Thermal Equilibrium'}

    In a steady-state, the gravitational radiation pumps energy into the r-mode at a rate given by
    
\begin{equation}
  W_d =  (1/3)\Omega J_c = -2\widetilde{E}/\tau_{GR}
\label{eq28}
\end{equation} 
\noindent
In a thermal steady-state, all of this energy must be dissipated in the star. Some fraction $L_\nu$ of this heat will be lost from the star due to neutrino emission and the rest $L_\gamma$ will be radiated at the surface. It should be mentioned that the thermal steady state is not an assumption but a rigorous result when the mode is saturated, and in particular it is independent of the cooling mechanism \cite{Alford2014}. We further assume that all of the energy emitted from the star during quiescence is due to the r-mode dissipation inside the star which implies $W_d=L_\nu+L_\gamma$ where $L_\nu$ and $L_\gamma$ are the neutrino luminosity and the thermal photon luminosity at the surface of the star, respectively.

\begin{table}[htbp]
\centering
\caption{Spin frequencies and core temperatures (measurements and upper limits) of observed Low Mass X-ray Binaries (LMXBs) and Millisecond Radio Pulsars (MSRPs) \cite{Ha12}.}
\begin{tabular}{||c|c|c||}
\hline
\hline
 Source&$ \nu$ (Hz)&$ T_{core} (10^8 K) $ \\ 
\hline
 Aql X-1&$550$&$1.08$ \\ 
\hline
 4U 1608-52&$620$&$4.55$ \\ 
\hline
 KS 1731-260&$526$&$0.42$ \\ 
\hline
 MXB 1659-298&$556$&$0.31$ \\ 
\hline
 SAX J1748.9-2021&$442$&$0.89$ \\
\hline 
 IGR 00291+5934&$599$&$0.54$ \\ 
\hline
 SAX J1808.4-3658&$401$&$<0.11$ \\ 
\hline
 XTE J1751-305&$435$&$<0.54$ \\ 
\hline
 XTE J0929-314&$185$&$<0.26$ \\ 
\hline
 XTE J1807-294&$190$&$<0.27$ \\ 
\hline
 XTE J1814-338&$314$&$<0.51$ \\ 
\hline
 EXO 0748-676&$552$&$1.58$ \\ 
\hline  
 HETE J1900.1-2455&$377$&$<0.33$ \\ 
\hline  
 IGR J17191-2821&$294$&$<0.60$ \\ 
\hline  
 IGR J17511-3057&$245$&$<1.10$ \\ 
\hline  
 SAX J1750.8-2900&$601$&$3.38$ \\ 
\hline
 NGC 6440 X-2&$205$&$<0.12$ \\ 
\hline
 Swift J1756-2508&$182$&$<0.78$ \\ 
\hline
 Swift J1749.4-2807&$518$&$<1.61$ \\ 
\hline
 J2124-3358&$203$&$<0.17$ \\ 
\hline 
 J0030+0451&$205$&$<0.70$ \\ 
\hline 
\hline
\end{tabular} 
\label{table4}
\end{table}
\noindent 
     
\begin{figure}[t]
\vspace{0.0cm}
\eject\centerline{\epsfig{file=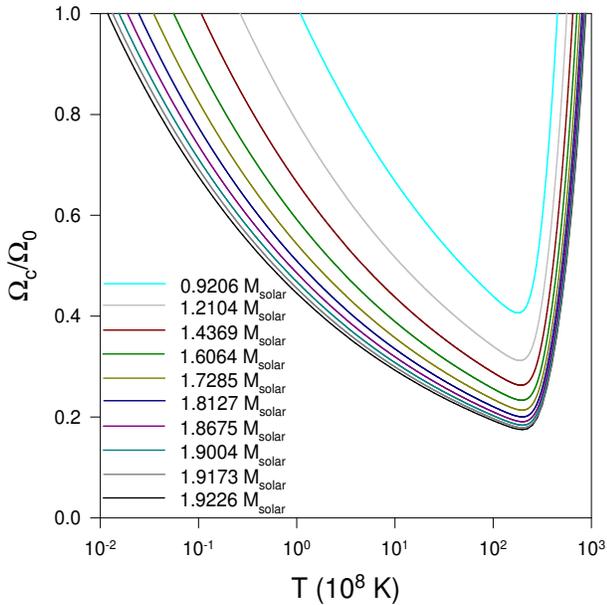,height=8cm,width=8cm}}
\caption{(Color online) Plots of reduced critical angular frequency with temperature for different masses of neutron stars.} 
\label{fig2}
\vspace{0.0cm}
\end{figure}
     
\noindent
\section{Results and discussion}
\label{Section 6}

    Present calculations have been performed using $\beta$-equilibrated neutron star matter obtained from density dependent M3Y effective interaction using saturation baryonic number density $\rho_0=0.1533$ fm$^{-3}$, saturation energy per baryon $\epsilon_0=-15.26\pm0.52$ MeV and the index of density dependence $n=\frac{2}{3}$  \cite{BCS08,CBS09}. The amplitude $\alpha_r$ can be fixed from `spin equilibrium' or `thermal equilibrium' as described in the preceding section.  
     
\noindent     
\subsection{Fiducial time scales} 
\label{Section 5}   
 
    The plots of the fiducial time scales with the gravitational masses of neutron stars using the DDM3Y EoS are shown in Fig.-\ref{fig1}. It is found that the gravitational radiation time scale falls rapidly with increasing mass. At the junction the shear viscous damping time scales increase approximately linearly while the shear viscous damping time scales in the core show opposite trend. We have also found that the contributions from the neutron-neutron scattering are more than that for the electron-electron scattering. However, the bulk viscous damping time scale remains almost constant with respect to NS masses. 

\noindent     
\subsection{r-mode instability window and Spin-down} 
\label{Section 5}
  
    The temperature $T$ dependence of the critical angular velocity $\Omega_c$ of the r-mode for the predominant $(l=2)$ mode can be obtained from Eq.(\ref{eq23}) by solving Eq.(\ref{eq24}) by knowing the fiducial gravitational radiation and viscous time scales. These results for critical angular velocity $\Omega_c$ in units of $\Omega_0$ have been plotted in Fig.-\ref{fig2} as a function of temperature up to $10^{11}$ K for several neutron star masses. 
 
\begin{figure}[t]
\vspace{0.0cm}
\eject\centerline{\epsfig{file=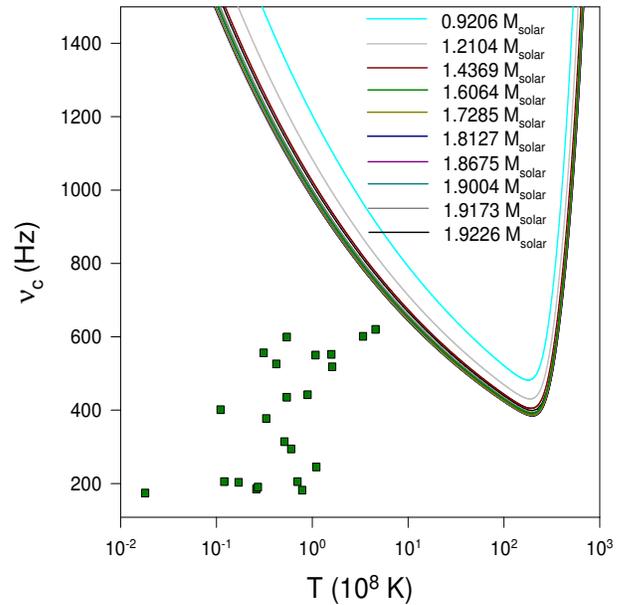,height=8cm,width=8cm}}
\caption{(Color online) Plots of critical frequency with temperature for different masses of neutron stars. The square dots represent observational data \cite{Ha12} of Table I. } 
\label{fig3}
\vspace{0.0cm}
\end{figure}
    
    The spin frequencies and core temperatures (measurements and upper limits) of the observed Millisecond Radio Pulsars (MSRPs) and Low Mass X-ray Binaries (LMXBs) \cite{Ha12} have been tabulated in Table-I and plotted in Fig.-\ref{fig3} in the present theoretical critical frequency versus temperature plot in order to compare their positions with respect to the instability window. It is interesting to find that the present $\beta$-equilibrated neutron star matter EoS for the core with a rigid crust, all of the observed neutron stars lie in the stable r-mode region which is consistent with the absence of observed gravitational radiation caused by r-mode perturbation.
    
\begin{figure}[t]
\vspace{0.0cm}
\eject\centerline{\epsfig{file=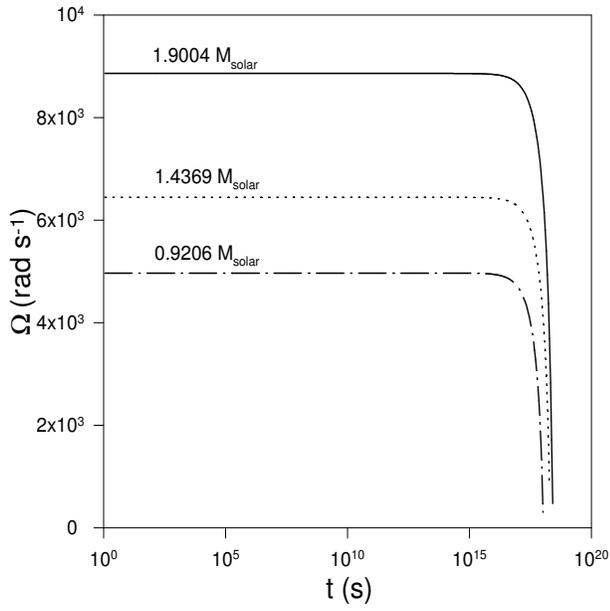,height=8cm,width=8cm}}
\caption{Plots of the evolution of angular frequencies with respect to time.} 
\label{fig4}
\vspace{1.46cm}
\end{figure}

\begin{figure}[b]
\vspace{0.0cm}
\eject\centerline{\epsfig{file=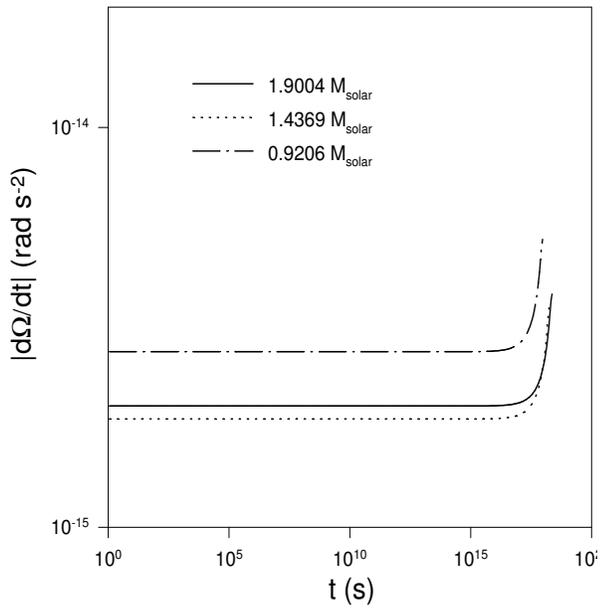,height=8cm,width=8cm}}
\caption{Plots of the evolution of spin-down rates with respect to time.} 
\label{fig5}
\vspace{0.0cm}
\end{figure}

    In Fig.-\ref{fig4} and Fig.-\ref{fig5}, the angular frequency and its rate of change have been plotted as functions of time for rotating NSs of three different masses, respectively. To solve Eq.(\ref{eq7}), the initial angular velocity has been fixed at the Keplerian angular velocity $\Omega_K\approx\frac{2}{3}\Omega_0$, since neutron star can never exceed this mass shedding limit. The upper limit of the r-mode amplitude $\alpha_r$ has been calculated from the quiescent luminosity of some particular neutron stars under the assumption that the heat energy produced due to emission of GW is radiated out through conventional cooling process. This is a good approximation in the cases of not too heavy mass NSs in which the direct URCA is not a possible cooling mechanism meaning that $L_\nu=0$. The amplitude of r-mode $\alpha_r$ has been estimated from `thermal equilibrium' using Eq.(\ref{eq28}) and $L_\gamma=4\pi R^2 \sigma T_{eff}^4$ where $\sigma$ is the Stefan's constant and $T_{eff}$ is the effective surface temperature of the star. The red-shifted effective surface temperature $T_{eff}^\infty$ has been chosen to be 100 eV which corresponds to $T_{eff}=T_{eff}^\infty/\sqrt{1-\frac{2GM}{Rc^2}}$ \cite{Beznogov2016}. The spin down due to gravitational wave emission (which continues until the critical angular frequency $\Omega_c$ is reached) is represented in Fig.-\ref{fig4} while Fig.-\ref{fig5} represents its rate. It is important to mention that r-mode amplitude $\alpha_r$ is minimum at about 1.44 M$_{\odot}$, effect of which is reflected in Fig.-\ref{fig5} implying that masses below it are more vulnerable to r-mode perturbations. 

\noindent     
\subsection{Gravitational Wave Amplitudes} 
\label{Section 5}
 
    When a neutron star accretes mass from its companion, angular momentum gets transferred and consequently its angular velocity increases and exceeds its critical value $\Omega_c$. At this point the neutron star starts emitting gravitational wave (GW). Due to CFS mechanism, GW emission pumps up the r-mode perturbation causing increase in amplitude $\alpha_r$ till it saturates. The saturation value can be estimated either from `spin equilibrium' or from `thermal equilibrium' described earlier. The emitted GW carries with it the angular momentum and energy and the star spins down to the region of stability. In the present calculations we complement the investigations of \cite{Mu18} by estimating the intensity of the GW emitted by the NSs.
    
\newpage

\begin{figure}[t]
\vspace{0.0cm}
\eject\centerline{\epsfig{file=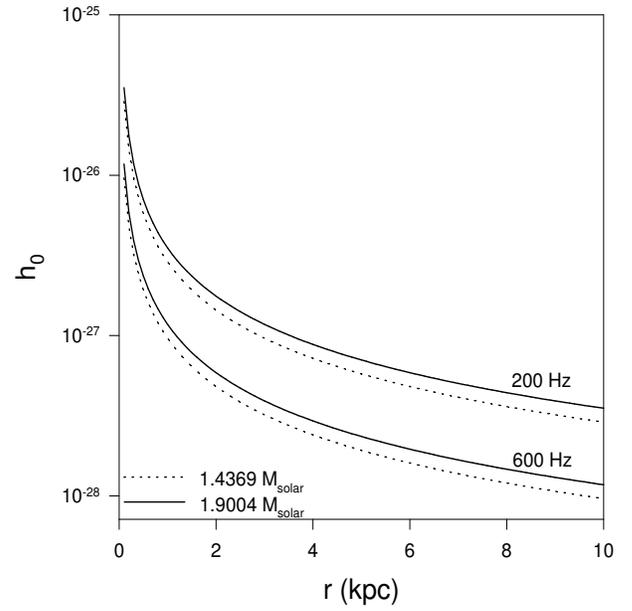,height=8cm,width=8cm}}
\caption{Plots of strain tensor amplitude $h_0$ as a function of distance.} 
\label{fig6}
\vspace{1.46cm}
\end{figure}

\begin{figure}[b]
\vspace{0.0cm}
\eject\centerline{\epsfig{file=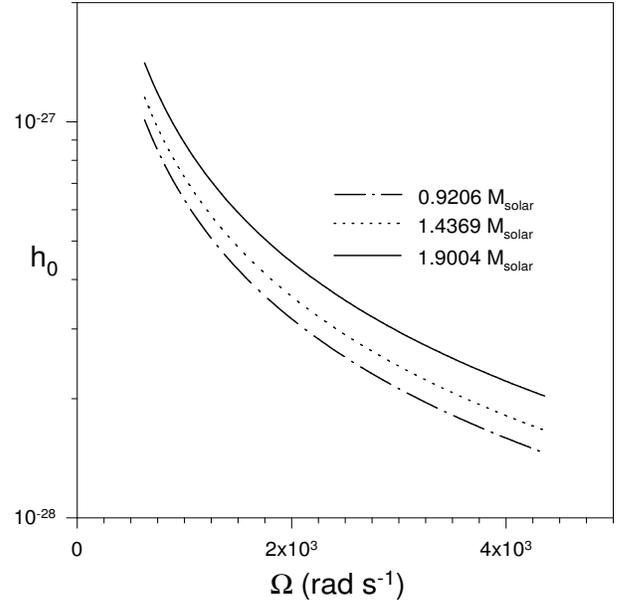,height=8cm,width=8cm}}
\caption{Plots of the amplitude of strain tensor $h_0$ as a function of reduced angular frequency from `thermal equilibrium'.} 
\label{fig7}
\vspace{0.0cm}
\end{figure}  
This is expressed in terms of the amplitude of the strain tensor, $h_0$. The amplitude of the strain tensor, $h_0$ is related to the $r$-mode amplitude $\alpha$ by \cite{Owen2009,Owen2010} 
     
\begin{eqnarray}
h_0=\sqrt{\frac{8\pi}{5}} \frac{G}{c^5} \frac{1}{r} \alpha_r \omega^3 M R^3\widetilde{J} , \nonumber
\label{eq99}
\end{eqnarray}
where $\omega$ is the r-mode angular frequency, which is related to the angular velocity of the star $\Omega$ by the relation $\omega=$.

\begin{figure}[t]
\vspace{0.0cm}
\eject\centerline{\epsfig{file=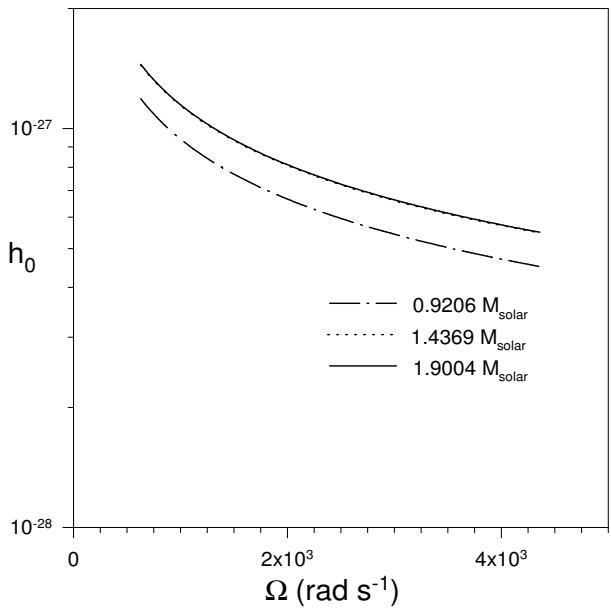,height=8cm,width=8cm}}
\caption{Plots of the amplitude of strain tensor $h_0$ as a function of reduced angular frequency from `spin equilibrium'.} 
\label{fig8}
\vspace{0.0cm}
\end{figure}

\noindent
This is expressed in terms of the amplitude of the strain tensor, $h_0$. The amplitude of the strain tensor, $h_0$ is related to the $r$-mode amplitude $\alpha$ by \cite{Owen2009,Owen2010} 
     
\begin{eqnarray}
h_0=\sqrt{\frac{8\pi}{5}} \frac{G}{c^5} \frac{1}{r} \alpha_r \omega^3 M R^3\widetilde{J} ,
\label{eq29}
\end{eqnarray}
where $\omega$ is the r-mode angular frequency, which is related to the angular velocity of the star $\Omega$ by the relation $\omega=-\frac{\left(l-1\right)\left(l+2\right)}{l+1} \Omega$. 
    
    In Fig.-\ref{fig6}, the amplitude of the strain tensor $h_0$ has been plotted as a function of distance for two different NS masses rotating with two fixed frequencies of 200 Hz and 600 Hz. Here also $\alpha_r$ has been estimated from `thermal equilibrium' and the red-shifted effective surface temperature $T_{eff}^\infty = 100$ eV has been used. For same $T_{eff}^\infty$, effective surface temperature $T_{eff}$ for higher mass would be higher which explains the trend resulting from higher $\alpha_r$ and correspondingly higher amplitude of the strain tensor $h_0$. The trend will be reversed if instead of $T_{eff}^\infty$, $T_{eff}$ is chosen to be the same. In Fig.-\ref{fig7}, the amplitude of the strain tensor $h_0$ has been plotted as a function of evolving angular velocity of the star $\Omega$ in units of rad s$^{-1}$ resulting from the GW emission for three different NS masses. On the contrary, Fig.-\ref{fig8}, is same but $\alpha_r$ has been estimated from `spin equilibrium' using Eq.(\ref{eq25}) and experimental data from the source IGR J00291 of Table-2 of Ref.\cite{Mahmoodifar2013}. The results for NS mass 1.9004 M$_\odot$ is very close but slightly higher than that for mass 1.4369 M$_\odot$ which is almost indistinguishable in Fig.-\ref{fig8}. The $\alpha_r$ estimated from `spin equilibrium' is higher than that from `thermal equilibrium' which corresponds to larger $h_0$ as can be seen in Fig.-\ref{fig8}.    

\noindent
\section{Summary and conclusions}
\label{Section 7}

    In the present work we have studied the r-mode instability of slowly rotating neutron stars with rigid crusts using the EoSs Feynman-Metropolis-Teller (FMT) \cite{FMT49}, Baym-Pethick-Sutherland (BPS) \cite{BPS71} and Baym-Bethe-Pethick (BBP) \cite{BBP71} up to number density of 0.0582 fm$^{-3}$ covering the outer crustal region and $\beta$-equilibrated neutron star matter beyond. This EoS provides good descriptions for proton, $\alpha$ and cluster radioactivities, elastic and inelastic scattering, symmetric and isospin asymmetric nuclear matter and neutron star masses and radii \cite{BCM14}. We have calculated the fiducial time scales for gravitational radiation, bulk viscosity, shear viscosity in the core and at the junction. It is observed that the gravitational radiation and shear viscosity in the core time scales decrease with increasing neutron star mass, the shear viscous damping time scales at the junction exhibit an approximate linear increase while bulk viscous time scale remains almost constant. Further, we have studied the temperature dependence of the critical angular frequency for different neutron star masses. It is observed that the majority of the neutron stars do not lie in the r-mode instability region. This fact is highlighted in Fig.-\ref{fig3} where the spin frequencies and core temperatures of observed Low Mass X-ray Binaries and Millisecond Radio Pulsars \cite{Ha12} always lie below the region of r-mode instability. The implication is that for neutron stars rotating with frequencies greater than their corresponding critical frequencies will emit GWs leading to unstable r-modes. The angular velocity of a neutron star can never exceed the Keplerian angular velocity $\approx \frac{2}{3}\Omega_0$ and thus, there is a critical temperature corresponding to $\Omega_c/\Omega_0=2/3$. It is important to mention the straight line at $\Omega_c/\Omega_0=2/3$ in Fig.-\ref{fig2} intersects each curve at two points providing two values of the critical temperature $T_c$. For temperatures below lower $T_c$ any perturbation is suppressed due to the dominance of shear viscosity while for temperatures above higher $T_c$ any perturbation is suppressed due to the dominance of bulk viscosity. The conclusion is that massive hot neutron stars are more susceptible to r-mode instability through gravitational radiation. We have also calculated the spin down rates and angular frequency evolution of the neutron stars through gravitational emission due to r-mode perturbation. Finally, the saturation value of r-mode amplitude $\alpha_r$ has been estimated from two different considerations of spin and thermal equilibria. Subsequently, the amplitude of the strain tensor $h_0$ has been calculated as a function of $\Omega$ and observer distance.


\end{document}